\documentclass[jap,graphicx, reprint]{revtex4-1} 
\usepackage{amsfonts, amsmath, graphicx}
\usepackage[version=3]{mhchem}

\begin{document}
\title{Effective Hamiltonian for surface states of \ce{Bi2Te3} nanocylinders with hexagonal warping} 
\author{Zhuo Bin Siu}
\affiliation{Computational Nanoelectronics and Nanodevices Laboratory, Electrical and Computer Engineering Department, National University of Singapore, Singapore} 
\author{Mansoor B. A. Jalil} 
\affiliation{Computational Nanoelectronics and Nanodevices Laboratory, Electrical and Computer Engineering Department, National University of Singapore, Singapore} 
\author{Seng Ghee Tan} 
\affiliation{Data Storage Institute, Agency for Science, Technology and Research (A*STAR), Singapore} 

\begin{abstract}
	The three-dimensional topological insulator \ce{Bi2Te3} differs from other topological insulators in the \ce{Bi2Se3} family in that the effective Hamiltonian of its surface states on a flat semi-infinite slab requires the addition of a cubic momentum hexagonal warping term in order to reproduce the experimentally measured constant energy contours. In this work, we derive the appropriate effective Hamiltonian for the surface states of a \ce{Bi2Te3} \textit{cylinder} incorporating the corresponding hexagonal warping terms in a cylindrical geometry. We show that at the energy range where the surface states dominate, the effective  Hamiltonian adequately reproduces the dispersion relation obtained from a full four-band Hamiltonian, which describe both the bulk and surface states. As an example application of our effective Hamiltonian, we study the transmission between two collinear \ce{Bi2Te3} cylinders magnetized in different directions perpendicular to their axes. We show that the hexagonal warping term results in a transmission profile between the cylinders which may be of utility in a multiple state magnetic memory bit. 
	
\end{abstract} 

\maketitle

\section{Introduction}
	The experimentally measured \cite{Nat460_1101, Sci325_178}  Fermi surfaces for the surface states of flat semi-infinite slabs of the three-dimensional topological insulator \ce{Bi2Te3} differs at moderate Fermi energies from the circular cross sections of Dirac cones predicted by the Dirac fermion Hamiltonian $H = v(\vec{p}\times\hat{z})\cdot\vec{\sigma}$ where $\hat{z}$ is the surface normal to the slabs. In order to reproduce the experimentally observed Fermi surfaces, Fu introduced a hexagonal warping term to the Hamiltonian based on the symmetries of the underlying crystal structure \cite{PRL103_266801}. The modified effective Hamiltonian for the surface states of the flat slab now reads as 
\begin{equation}
	H = v(\vec{p}\times\hat{z})\cdot\vec{\sigma} + \lambda \sigma_z (k_x^3-3k_xk_y^2).  \label{ham} 
\end{equation} 
	
	Liu \textit{et al} subsequently derived a four-band effective Hamiltonian \cite{PRB82_045122} which describes both the bulk as well as surface states near the Dirac point of the \ce{Bi2Se3} family of TIs, of which \ce{Bi2Te3} is a member. They noted that Fu's hexagonal warping term can also obtained from the model Hamiltonian. 
	
	We showed in an earlier work \cite{SciRep4_5062} that the additional hexagonal warping term in a flat \ce{Bi2Te3} slab leads to the distortion of the Fermi surface and the opening of a bandgap by an \textit{in-plane} magnetization, both of which would not have occured in the absence of the hexagonal warping term. These and the displacement of the Fermi surfaces by the in-plane magnetization lead to a complex profile of the transmission from an unmagnetized source segment to a magnetized drain segment, as a function of the magnetization direction and the electron energy.
	
	Besides flat slabs and flakes, another common morphology for experimentally grown TI nanostructures is the nanowire \cite{SciRep3_1212, SciRep3_1564, APL103_193107}, which can approximately be described as a cylinder. The effective Hamiltonian for the surface states of a TI cylinder \textit{without} the hexagonal warping term has been derived to be \cite{PRB79_245331,PRL105_206601,PRL105_136403} 
\begin{equation}
		H = v(p_\phi\sigma_z - p_z\sigma_\phi).  \label{Hcy0} 
\end{equation} 
	
	The surface states of a TI cylinder differ from those of the flat TI slab in two important aspects. First, the circumference of the cylinder imposes periodic boundary conditions on the surface states along the azimuthal direction. Second, the curvature along the circumference of the cylinder leads to a position dependent normal vector to the surface. This in turn leads to a position dependent spin orbit interaction field, which is reflected by the presence of the $\sigma_\phi\equiv (\cos(\phi)\sigma_x+\sin(\phi)\sigma_y)$ term in the Hamiltonian. 
		
	The above observations motivate us to analyze the combined effects of the  of the hexagonal warping term together with a cylindrical geometry in \ce{Bi2Te3} nanowires. Moreover, whilst the effective Hamiltonian for the surface states of a cylindrical TI nanowire has already been derived,\cite{PRB84_195406} the hexagonal warping terms have been neglected in the derivation. Thus in this work, we derive the effective Hamiltonian for the surface states of a \ce{Bi2Te3} nanowire which now includes the hexagonal warping term starting from the four-band Hamiltonian of Liu \textit{et al}. To test the accuracy of the effective surface state Hamiltonian, we compare the bandstructure from the Hamiltonian with that from the full four-band Hamiltonian, and show reasonable matching between the two bandstructures. 
	
	As an example applicatin of the surface state Hamiltonian, we study the transmission between two \ce{Bi2Te3} magnetized in different directions perpendicular to the cylinder axis.  The hexagonal warping term leads to a dependence of the energy-$k_z$ relation on the magnetization direction in the plane of the cylinder circumference. This in turn results in a sawtooth transmission profile in the transmission between two \ce{Bi2Te3} cylinders magnetized in different directions, a feature which may be of utility in achieving multiple-state magnetic memory bits.

\section{Derivation of Effective Hamiltonian}  
We take as our starting point the model four-band Hamiltonian in Eqs. 16 and 17 of Ref. [\onlinecite{PRB82_045122}],
\begin{eqnarray}
	H_{(4B)} &=& (M_0 + M_1 k_z^2 + M_2 k_\parallel^2)\Gamma_5 + B\Gamma_4k_z   \nonumber \\ 
	&&  + A(\Gamma_1k_y-\Gamma_2k_x) + R_1\Gamma_3(k_x^3-3k_xk_y^2) \nonumber \\ 
	&& + R_2\Gamma_4(3k_x^2k_y-k_y^3) \label{HCa} 
\end{eqnarray}
where
\begin{eqnarray*}
	&& \Gamma_1 = \sigma_x\otimes\tau_1,	\Gamma_2 = \sigma_y\otimes\tau_1, \\
	&& \Gamma_3 = \sigma_z\otimes\tau_1, \Gamma_4 = \mathbb{I}_\sigma\otimes\tau_2, 	\Gamma_5 = \mathbb{I}_\sigma\otimes\tau_3, 
\end{eqnarray*}
and $k_\parallel^2 \equiv k_x^2+k_y^2$. The $\tau$'s can be interpreted as describing the orbital degree of freedom and the $\sigma$ terms the real spins.  This Hamiltonian can be derived based on the symmetries exhibited by the underlying crystal structure of the \ce{Bi2Se3} family of TIs, and describes both the bulk as well as surface states.  (The subscript (4B) stands for ``4-band''. )

We first derive the two-band effective Hamiltonian for the surface states of a TI (\ce{Bi2Te3}) cylinder with its axis parallel to the $z$ direction without the $R_1$ and $R_2$ terms (i.e. without the hexagonal warping), and incorporate the effect of curvature by an approach modified from Imura \textit{et al} \cite{PRB84_195406}. This would yield a Hamiltonian compatible to the usual Dirac Hamiltonian of Eq. \eqref{Hcy0}.  The general idea is to first split the four-band Hamiltonian $H_{(4B)}$, which describes both the bulk and surface states, into two parts which we call the perpendicular Hamiltonian $H_{(4B),\perp}$ and parallel Hamiltonian $H_{(4B),\parallel}$ for short. The first part  contains momentum operators perpendicular to the surface, and the second part momentum operators parallel to the surface. 

We know that the surface states exist near the zero-energy Dirac point where the energy contribution due to the perpendicular Hamiltonian is minimal. In other words, the energy of these states should come only from the momentum of the particles moving on the surface. We hence find the zero energy ground states of the perpendicular Hamiltonian which are localized near the surface, and then treat the parallel Hamiltonian as a perturbation to the perpendicular Hamiltonian ground states. The ground states are degenerate due to the presence of the orbital and spin degrees of freedom. According to degenerate perturbation theory, the first order effects of a perturbation are given by projecting the perturbing Hamiltonian in the basis of the degenerate ground states. The parallel Hamiltonian projected into the basis of the degenerate ground states thus yields the effective Hamiltonian for the states localized near the surface.

 In detail, we first identify $k_y\rightarrow(-i\partial_y)$, $(-i\partial_y) = - i(\frac{\partial r}{\partial_y}\partial_r + \frac{\partial \phi}{\partial_y}\partial_\phi)$, etc., to rewrite Eq. \eqref{HCa} in cylindrical coordinates $(r, \phi)$. We treat the terms containing $r$ derivatives resulting from the $(M_0 - M_2\partial_r^2)\Gamma_5 + A(\Gamma_1 k_y-\Gamma_2 k_x)$ as our base Hamiltonian and the remaining terms as perturbations. 

\subsection{Zero energy eigenstates of $H_{(4B),\perp}$}
The perpendicular Hamiltonian in cylindrical coordinates reads as
\begin{eqnarray}
	\langle\vec{r}|H_{(4B),\perp} &=& (M_0 - M_2\partial_r^2)(\mathbb{I}_\sigma\otimes\tau_3) + A ( - (\sigma_\phi)(-i\partial_r) )\otimes\tau_1 \nonumber  \\
	&\equiv& M_\perp(\mathbb{I}_\sigma\otimes\tau_3) + A ( - (\sigma_\phi)(-i\partial_r) )\otimes\tau_1
	\label{H0Cy} 
\end{eqnarray}  
where, following Ref. [\onlinecite{PRB84_195406}], we dropped the term resulting from $M_2(\frac{1}{r})(-\partial_r)$. This is justifiable on the basis that the edge states we are seeking are confined to around $r=R$ where in the $R\rightarrow\infty$ limit the contributions of terms proportional to $\frac{1}{R}$ go to 0. As mentioned above, we look for eigenstates of  Eq. \eqref{H0Cy} with eigenenergy 0 and with spatial $r$ dependence of the form $\exp(\lambda(r-R))$.

It turns out to be physically more instructive to diagonalize the $\sigma$ part of Eq. \eqref{H0Cy} first. The $\sigma$ eigenstates $|\sigma_\pm\rangle$ can be  solved to be $|\sigma_\pm\rangle \equiv \frac{1}{\sqrt{2}} (1, \pm i \exp(i \phi))^\mathrm{T}$. Substituting the eigenstates into Eq. \eqref{H0Cy}, we then have an equation for the $\tau$ degree of freedom -- 
\begin{equation}
	\langle\vec{r}, \sigma_\pm|H_{(4B), \perp} = (M_\perp \tau_3  \pm i A\lambda \tau_1)\langle\vec{r},\sigma_\pm|.
	\label{H1Cy} 
\end{equation}

We now solve for the zero energy eigenstates of Eq. \eqref{H1Cy}.  Writing the (unnormalized) zero energy eigenstate of Eq. \eqref{H1Cy} as $(1, \alpha)^\mathrm{T}$ and solving for $\alpha$ yields the eigenstates $(1, \pm \frac{i M_\perp}{A\lambda})^\mathrm{T}$. Setting the secular equation for the eigenvalues of Eq. \eqref{H1Cy} to be 0 then yields $(M_\perp)^2 = (A\lambda)^2$.

By definition $M_\perp = M_0 - \lambda^2 M_2$. A condition for the existence of topological edge states is that $M_0$ and $M_2$ need to be of differing signs. It is conventional to choose $M_0$ to be negative, in which case $M_\perp$ is negative. As $M_\perp$ is negative and $A$ is positive, the only consistent solution for  $(M_\perp)^2 = (A\lambda)^2$ is to choose $M_\perp=-A\lambda$. The (normalized) zero-energy eigenstates of Eq. \eqref{H1Cy} are then
\[
		|\tau_\pm\rangle = \frac{1}{\sqrt{2}} \begin{pmatrix} 1 \\ \mp i \end{pmatrix}. 
\]

We note that the equation $M_\perp=-A\lambda$ actually expands out to $(M_0-\lambda^2M_2)=-A\lambda$ which is quadratic in $\lambda$ and hence yields two solutions which we denote as $\lambda_{(\pm)}$. Note that the bracketed $(\pm)$ in this subscript differs from the $\pm$ index which pertains to the spin ($\sigma$) eigenstate of Eq. \eqref{H0Cy}.
 Explicitly, we have 
\begin{equation}
	\lambda_{(\pm)} = \frac{A\pm\sqrt{A^2+4M_0M_2}}{2M_2}. \label{lambdaPM}
\end{equation}
 
Putting everything together, a zero-energy eigenstate of Eq. \eqref{H1Cy} which vanishes at the cylinder boundaries $r=R$ is given by
\begin{eqnarray*}
		&& \langle \vec{r}|\pm\rangle \\
		&\propto&  \frac{\exp(\lambda_{(+)} (r-R)) - \exp(\lambda_{(-)}(r-R))}{2} \begin{pmatrix} 1 \\ \mp i \\ \pm i\exp(i\phi) \\ \exp(i\phi) \end{pmatrix}. 
\end{eqnarray*}

To obtain the normalization constant, we evaluate the following integral in the limit of  $\lambda_{(\pm)}R \gg 1$, 
\begin{eqnarray*}
	&& \int^R_0 \ \mathrm{d}r\  2\pi r (\exp(\lambda_{(+)} (r-R))-\exp(\lambda_{(-)} (r-R)))^2\\
	&\approx& \frac{\pi R (\lambda_{(-)}-\lambda_{(+)})^2 }{\lambda_{(-)}\lambda_{(+)}(\lambda_{(+)}+\lambda_{(-)})}.
\end{eqnarray*}
(Ref. [\onlinecite{PRB84_195406}] writes of the $\lambda_{(\pm)} \ll R$ limit but we find that $\lambda_{(\pm)}R \gg 1$ is a better description of what we end up operationally doing.) The normalized zero-energy eigenstates thus read as
\begin{equation}
	\langle \vec{r}|\pm\rangle = \rho(r)\begin{pmatrix} 1 \\ \mp i \\ \pm i\exp(i\phi) \\ \exp(i\phi) \end{pmatrix} 
\end{equation}
where $\rho(r)  = \sqrt{\frac{\lambda_{(-)}\lambda_{(+)}(\lambda_{(+)}+\lambda_{(-)})}{\pi R (\lambda_{(-)}-\lambda_{(+)})^2 }}[\exp(\lambda_{(+)} (r-R))-\exp(\lambda_{(-)} (r-R))]$. 

\subsection{Projection of $H_{(4B),\parallel}$ into zero-energy $(2\times2)$ subspace of $H_{(4B),\perp}$}
We now proceed to derive the effective Hamiltonian on the surface state of the \ce{Bi2Te3} cylinder. We first consider $H_{(4B),0}=(M_0 + M_1 k_z^2 + M_2 k_\parallel^2)\Gamma_5+B\Gamma_4 k_z + A(\Gamma_1 k_y - \Gamma_2 k_x)$ [where the ``0"' in the subscript denotes the absence of the $R_1$ and $R_2$ hexagonal warping terms in Eq. \eqref{HCa}]. From this, we extract the Hamiltonian $H_{(4B),\parallel; 0}$ that corresponds to motion parallel to the surface of the cylinder [i.e., the terms which are not already included in $H_{(4B), \perp}$ in Eq. \eqref{H0Cy}]:
\[ 
	\langle\vec{r}|H_{(4B),\parallel; 0} = (M_1 k_z^2)\Gamma_5+B\Gamma_4 k_z - i\frac{A}{r}\sigma_r\otimes\tau_1\partial_\phi.
\]

We then project $H_{(4B),\parallel; 0}$ into the subspace spanned by $|\pm\rangle$, the zero-eigenstates of $H_{(4B),\perp}$. In other words, we treat $H_{(4B),\parallel;0}$ as a perturbation to $H_{(4B),\perp}$. An explicit calculation of $\int \mathrm{d}r\ r \langle\vec{r},\pm'|H_{(4B),\parallel; 0}|\vec{r},\pm\rangle$ for the four possible combinations of $\pm'$ and $\pm$ yields  
\begin{eqnarray*}
	\tilde{H}_{(2B);0}  &\equiv&   \begin{pmatrix} \langle  +|H_{(4B),\parallel;0}|+ \rangle &  \langle  +|H_{(4B),\parallel;0}|- \rangle \\  \langle -|H_{(4B),\parallel;0}|+\rangle &  \langle  -|H_{(4B),\parallel;0}|-\rangle \end{pmatrix} \\
	&=& \begin{pmatrix} B k_z & A p_\phi  \\
	 A p_\phi & - B k_z \end{pmatrix}.
\end{eqnarray*} 

The spin components of the matrix above are in the $| \sigma_\pm\rangle$ basis. They can be converted to the original spin basis of Eq. \eqref{HCa} through the unitary transformation $H_{(2B);0} = U\tilde{H}_{(2B);0}U^\dagger$ with 
\[
U = \sum_{\sigma=(\uparrow/\downarrow),\tilde{\sigma}=\sigma_{\pm}} |\sigma\rangle\langle\sigma|\tilde{\sigma}\rangle\langle\tilde{\sigma}|.
\]
This yields 
\begin{equation}
	H_{(2B);0} = A \left(p_\phi\sigma_z+\frac{1}{2R}\mathbb{I}_\sigma\right) - B k_z\sigma_\phi \label{H1ti},
\end{equation}
where $p_\phi\equiv(-i\partial_\phi)/R$. Eq. \eqref{H1ti} coincides with the usual Rashba form of the Dirac fermion Hamiltonian $v (\vec{\sigma}\times\vec{p})\cdot\hat{n}$ on the surface of the cylinder expanded out in cylindrical coordinates with the identification of $A=B/\hbar=v$ except for the presence of the additional $\frac{A}{2R}$ term. The latter has been interpreted as a manifestation of the spin Berry phase \cite{PRB84_195406} but can be ignored for most transport calculations as a constant shift in energy.

Finally, we consider the remaining terms in Eq. \eqref{HCa} that have not yet been dealt with, and which correspond to the warping terms, i.e. 
\[
	 H_{(4B),\parallel;R}=R_1\Gamma_3 (k_x^3-3k_xk_y^2) + R_2\Gamma_4(3k_x^2k_y-k_y^3). 
\]

These terms require a bit more subtlety to handle as they contain $r$ and its derivatives. We often see the momentum in the $x$ direction being written as $\hat{p}_x\simeq-i\partial_x$, so that the expectation value of the momentum at position $x$ gets written as $\psi(x)^*(-i\partial_x \psi(x))$. We should, however, recognize that $\psi(x)^*(-i\partial_x \psi(x))$ is actually $\langle \psi| \Big( |x\rangle\langle x|p_x\Big)|\psi\rangle$ where we are taking the expectation value of the \textit{non-Hermitian} operator $|x\rangle\langle x|p_x$ with respect to the state $|\psi\rangle$. This does not lead to any problems when $\psi(x)=\langle x|\psi\rangle = \exp(i k x)$ where $k$ is real, but yields a complex value where $k$ is complex. This contravenes the fact that an expectation value which corresponds to a physical measurement should yield a real value. The most natural way around this is to modify the expression for the momentum to $\mathrm{Re} (\psi^*(-i\partial_x \psi)) = \frac{1}{2} ( (i\partial_x\psi^*)\psi + \psi^*(-i\partial_x\psi)) = \frac{1}{2}\langle \psi| \Big( \{ |x\rangle\langle x|, p_x\} \Big) |\psi\rangle$. The preceding discussion is relevant for the case of a TI cylinder because the $r$ dependence of the $|\pm\rangle$ states take the form of $\exp(- r\lambda)$ which can be interpreted as a wave with a complex wavevector. Following our discussion,  the matrix elements of the projected $(2\times2)$ Hamiltonian $H_{(2B);R}$ which represents the warping terms acting on the surface states, are given by
\begin{eqnarray}
	 \left[H_{(2B);R}\right]_{\pm',\pm} &=& \frac{1}{2} \langle \pm'|\{ R_1 \Gamma_3 (p_x^3-3p_xp_y^2)\nonumber \\
	&& + R_2 \Gamma_4 (3p_x^2p_y-p_y^3), |\vec{r}\rangle\langle\vec{r}|\}|\pm\rangle.\nonumber
\end{eqnarray}
Evaluating the integrals, dropping terms containing $\exp(-\lambda R)$, and performing the unitary transformation to the real spin basis as before, gives the final expression
\begin{eqnarray}
	H_{(2B)} &\equiv& H_{(2B;0)} +H_{(2B;R)} \nonumber\\
	&=&-B k_z \sigma_\phi  + \frac{A}{R}\left(\sigma_z(-i\partial_\phi)+\mathbb{I}_\sigma\frac{1}{2}\right) \nonumber \\
	&-&\frac{3i\lambda_{(-)}\lambda_{(+)}}{2R} \{ \partial_\phi,(R_1 \sin(3\phi)\sigma_r - R_2 \cos(3\phi)\sigma_\phi)\}.\nonumber\\  \label{Heff}
\end{eqnarray}	
This effective Hamiltonian is the key result of this work. The steps outlined in the previous section can generally be applied to reduce a full 4-band Hamiltonian to a $(2\times2)$ surface state Hamiltonian. For instance, the surface state Hamiltonian for a flat TI slab with warping term in Eq. \eqref{ham}, as obtained by Fu, can be obtained from the full Hamiltonian in Eq. \eqref{HCa} by following the same procedure, but with the TI cylinder replaced by a TI crystal truncated perpendicular to the $z$ direction, as we briefly outline in the Appendix.

We then find that the parameters $v$ and $\lambda$ in Eq. \eqref{ham} correspond to $A$ and $R_1$ in Eq. \eqref{HCa}, respectively. The values of $A$ in both Refs. [\onlinecite{PRL103_266801}] and [\onlinecite{PRB82_045122}] match within 10\% ($2.55\ \mathrm{eV}$\AA  versus $2.87\ \mathrm{eV}$\AA, respectively)	 and we shall use the latter value for our numerical calculations. The values of $R_1$, however, differ by a factor of 5.5 ($250~\mathrm{eV}$\AA$^3$ versus $45.02~\mathrm{eV}$\AA$^3$) between the two works. We assume the value of  $R_1$ to be 5 times the latter value in our numerical calculations. This value is chosen as it is intermediate between the values quoted in the two works, but closer to towards that in Ref. [\onlinecite{PRL103_266801}] which was obtained by comparison to actual experimental data, and hence may be deemed to be more reliable.    

\section{Comparison between two-band and four-band Hamiltonians}
In this section, we compare how well the effective two-band Hamiltonian of Eq. \eqref{Heff} for the surface states replicates the surface states obtained via the full four-band Hamiltonian of Eq. \eqref{HCa}. To accomplish this, we write the spatial parts of the latter in the basis of the normalized eigensolutions to the two-dimensional cylindrical coordinate Laplacian equation which vanish at $r=R$, thereby satisfying the boundary condition that the eigenstate wavefunctions go to 0 at the surface of the cylinder.  These eigensolutions $|m,l\rangle$ are indexed by the quantum numbers $m$ and $l$ which carry the usual quantum mechanical interpretations of the $z$ and total angular momentum respectively, and are given by  
\[
	\langle \vec{r}|m, l\rangle = \frac{\exp(i m\phi) J_m\left(\frac{r}{R} j_{m,l}\right)j_{m,l}^2}{\sqrt{\pi}R^3J_{m+1}(j_{m,l})}
\]
where $J_{m}$ is the Bessel function of the first kind of order $m$, and $j_{m,l}$ its $l$th zero. (A theorem in elementary quantum mechanics states that the eigenstates for a Hermitian Hamiltonian form a complete basis set, so we do not need to consider states proportional to $\exp(i m\phi)J_n(\frac r{R} j_{n,l})$ for $m\neq n$.)  The resulting matrix is then diagonalized numerically to yield the energy eigenvalues.

There is a technical caveat that has to be considered in evaluating the matrix elements involving the third order derivatives in $r$, which occur in the $R_1$ and $R_2$ hexagonal warping terms. For example, the terms in the Hamiltonian containing $R_1$ terms acting on arbitrary states $\langle \Psi|$ and $|\Phi \rangle$ can be most succinctly be expressed by 
\begin{eqnarray*} 
	&& R_1 \Psi^\dagger \Gamma_3(k_x^3-3k_xk_y^2) \Phi \\
	=&&R_1 \mathrm{Im} \Psi^\dagger\Gamma_3\Big[\cos3\phi\Big(-\frac{3 \partial_\phi^2}{r^3}-\frac{3 \partial_r}{2 r^2}+\frac{3 \partial_r\partial_\phi^2}{2 r^2}+\frac{3 \partial_r^2}{2 r}\\
	&-&\frac{1}{2} \partial_r^3 \Big)+\sin3\phi\Big(\frac{4 \partial_\phi}{r^3}-\frac{\partial_\phi^3}{2 r^3}-\frac{9 \partial_r\partial_\phi}{2 r^2}+\frac{3 \partial_r^2\partial_\phi}{2 r}\Big)\Big]\Phi.
\end{eqnarray*}
One might then naively evaluate the $\partial^3_r$ matrix elements $\langle m',l'\partial^3_r |m,l\rangle$ as 
\begin{equation}
	\int^R_0 \mathrm{d}r\  r J_{m'}(r/R j_{m',l'})\partial^3_r J_m(r/R j_{m,l}). \label{r30}
\end{equation}
It turns out, however, that such a na\"{i}ve calculation would not yield a Hermitian Hamiltonian. A more proper interpretation of the matrix element should be  
\begin{equation}
\int^\infty_0 \mathrm{d}r\ r J_{m'}(r/R j_{m',l'})\partial^3_r (\Theta(R-r)J_m(r/R j_{m,l})),\label{r31}
\end{equation} 
where $\Theta$ is the Heaviside step function. Compared to Eq. \eqref{r30}, Eq. \eqref{r31} has extra terms after integration due to the Dirac delta functions that arise from differentiating the step function. \footnote{These extra terms are also present in the evaluation of integrals involving the second $r$ derivatives but evaluate to 0 as they involve a product of a $J_m(r/R j_{m,l})$ and the derivative of another one at $r=R$.} For matrix elements involving third $r$ derivatives, the extra terms include products of derivatives of Bessel functions which do not evaluate to 0 at $r=R$, and have to be accounted for.

	The calculation of the dispersion relations with the full four-band Hamiltonian enables us to compare how well our effective two-band Hamiltonian approximates the surface states of the four-band Hamiltonian. The calculation also establishes the energy and wavevector ranges above which bulk states start to emerge, and would invalidate any transport calculations based on the two-band Hamiltonian, which considers only the surface states. 
	
\begin{figure}[ht!]
\centering
\includegraphics[scale=0.5]{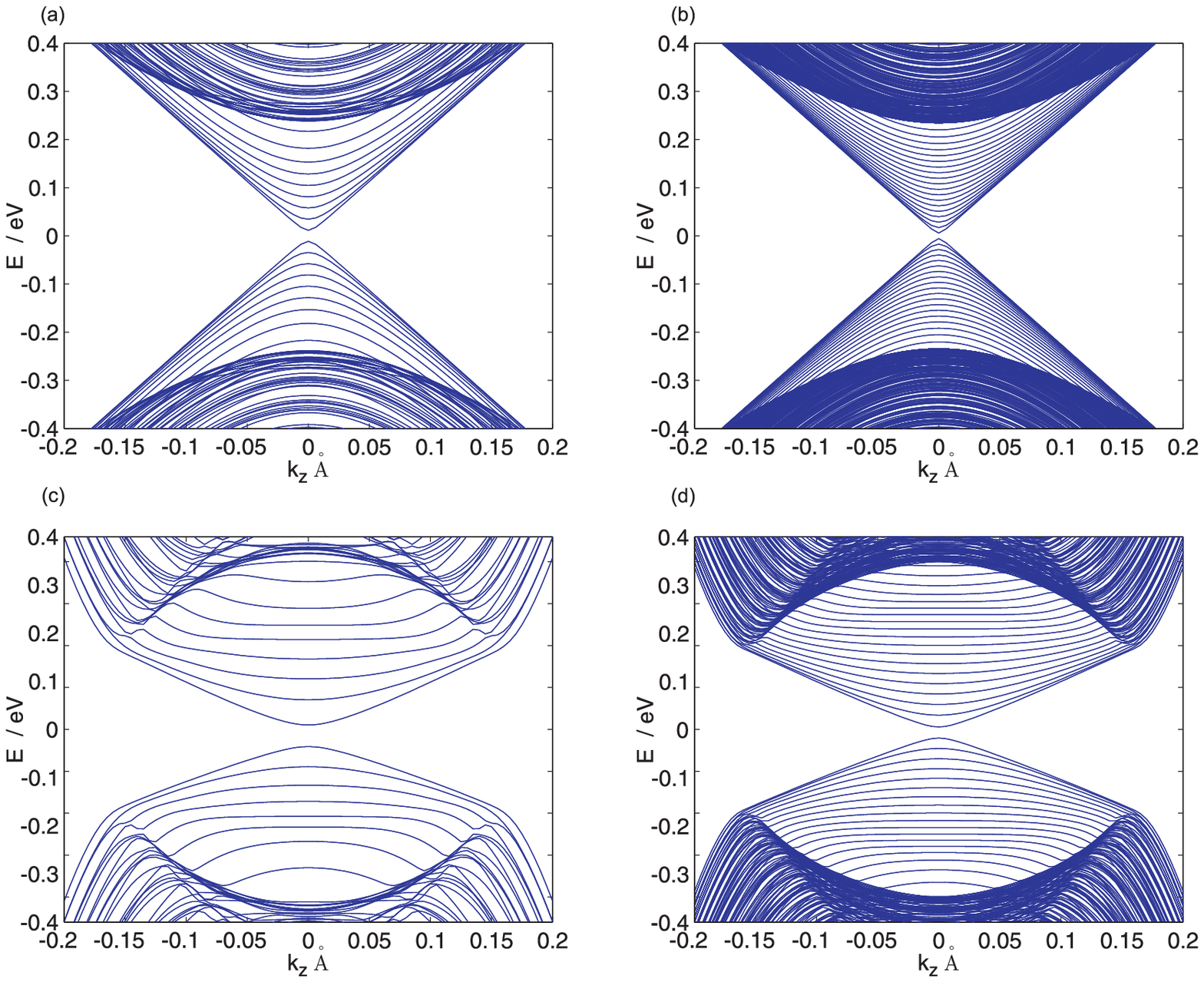}
\caption{  Dispersion relations for the edge states and some bulk states for  (a), (b) \ce{Bi2Se3} and (c), (d) \ce{Bi2Te3} for TI nanocylinders of radii  (a),(c) $15~\mathrm{nm}$  and (b),(d) $30~\mathrm{nm}$. } 
\label{gcombEK1}
\end{figure}

	Fig. \ref{gcombEK1} shows the dispersion relations calculated for \ce{Bi2Se3} and \ce{Bi2Te3}  using the four-band Hamiltonian parameters given in Ref. [\onlinecite{PRB82_045122}] for TI nanocylinders except for the $R_1$ parameter for $\ce{Bi2Te3}$, for which we assume a value 5 times that quoted in the paper as discussed in the previous section. 
	
	As the radius of the cylinders increases, the energy spacings between the bulk and edge bands decrease. In \ce{Bi2Se3} the edge states are more cleanly separated in energy from the bulk states compared to \ce{Bi2Te3} where some of the higher energy edge states are found at energy ranges where bulk states are present as well. Somewhat surprisingly, the energy values (and for \ce{Bi2Te3}, the wavevectors) at which the bulk states emerge and render the two-band effective Hamiltonian for surface states inappropriate, is only very weakly dependent on the cylinder radii. The results indicate that the bulk states may be ignored up to around $0.25\ \mathrm{eV}$ in \ce{Bi2Se3} and $0.1\ \mathrm{eV}$ in \ce{Bi2Te3} cylinders, for the range of cylinder radii considered. 
 	
\begin{figure}[ht!]
\centering
\includegraphics[scale=0.55]{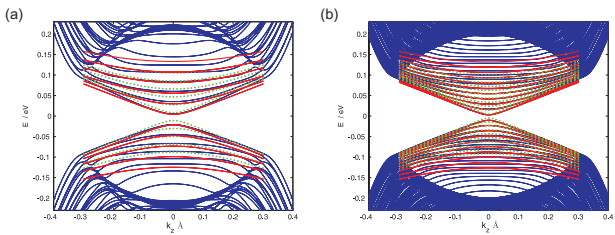}
\caption{  The dispersion relations for a TI cylinder of radius (a) $15\ \mathrm{nm}$ and (b) $30\ \mathrm{nm}$ calculated using the full four-band Hamiltonian (blue lines), the two-band effective Hamiltonian (red lines) and the two-band Hamiltonian without the hexagonal warping terms (green dots) .} 
\label{gefEkComp}
\end{figure}		
 
Next, we compare the dispersion relations for \ce{Bi2Te3} obtained from the two-band and four-band Hamiltonians. Fig. \ref{gefEkComp} shows the dispersion relations for two TI cylinders of different radii calculated using the full and the effective Hamiltonian, as well as the effective Hamiltonian without the hexagonal warping term $H_{(2B);0} = A \left(p_\phi\sigma_z+\frac{1}{2R}\mathbb{I}_\sigma\right) - B k_z\sigma_\phi$. The depicted edge states are two-fold degenerate, i.e., each distinct line in the figure represents two states.  In both panels of the figure, there is evidently a significant mismatch between the dispersion relations of the two-band and four-band Hamiltonians if the hexagonal warping terms of the two-band Hamiltonians were neglected (green dots). On the other hand, the dispersion relations obtained by the effective surface state Hamiltonian of Eq. \eqref{Heff} show a reasonably close fit with those obtained by the full Hamiltonian of Eq. \eqref{HCa} at small $k_z$ and at energy values lower than the threshold at which the bulk bands emerge. At larger values of $k_z$ below $0.1\ \mathrm{eV}$, the effective Hamiltonian tends to underestimate the eigenenergy. However the qualitative trend that the bottom-most pairs of degenerate particle states tend to cluster together at large $k_z$ is reproduced by the effective Hamiltonian. The agreement between the effective and full Hamiltonian $E-k$ curves improves with increasing cylinder radius. This is to be expected as the effective Hamiltonian was derived under the $R\rightarrow\infty$ limit.

\section{Dispersion relations and magnetoresistance} 
Having now established that our effective two-band Hamiltonian is an adequate approximation for the full four-band Hamiltonian for a \ce{Bi2Te3} nanowire, we now use the former to study a \ce{Bi2Te3} cylinder with an magnetization. We use the numerical value of $50\ \mathrm{nm}$ for the cylinder radius in our calculations. This choice is motivated by two reasons. The results of the previous section suggests a reasonably good match between  the eigenstates calculated by the four-band Hamiltonian and the two-band effective Hamiltonian can be obtained at $50\ \mathrm{nm}$ radius.  This radius value also falls within the range reported for experimentally fabricated \ce{Bi2Te3} nanowires \cite{SciRep3_1212,  SciRep3_1564}.

\begin{figure}[ht!]
\centering
\includegraphics[scale=0.67]{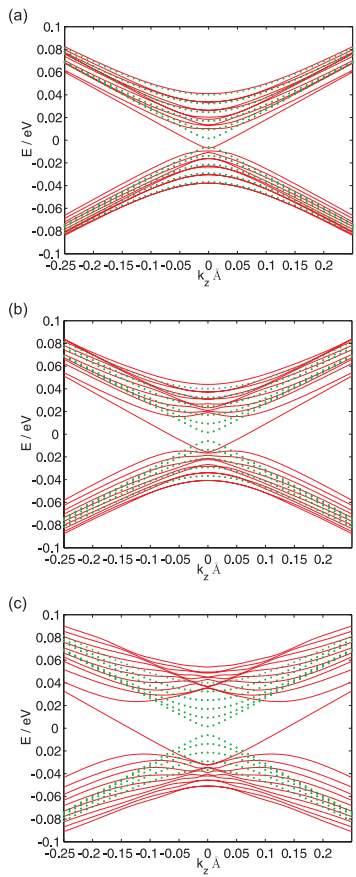}
\caption{   The dispersion relations for an magnetization applied in the $x$ direction for a \ce{Bi2Te3} nanowire of radius $500$ \AA, and (a) $M_x = 0.01\ \mathrm{eV}\hbar^{-1}$, (b) $M_x = 0.02\ \mathrm{eV}\hbar^{-1}$ and (c) $M_x = 0.04\ \mathrm{eV}\hbar^{-1}$ respectively plotted in solid lines, and the dispersion relations without the magnetizations plotted in dotted lines for comparison. } 
\label{gT2muXcombEK}
\end{figure}		

The panels of Fig. \ref{gT2muXcombEK} shows the dispersion relations for increasing magnetizations in the $x$ direction. Similar to the TI nanowires studied in Ref. [\onlinecite{JAP117_17C749}],  the energies of the particle states at $k_z=0$ increase with increasing magnetizations, and the low energy particle bands become convex at small $|k_z|$ at large magnetizations. 

\begin{figure}[ht!]
\centering
\includegraphics[scale=0.48]{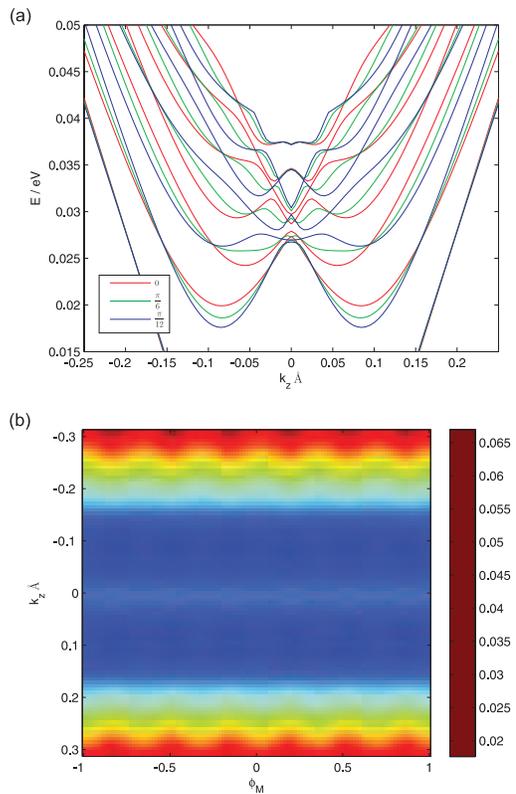}
\caption{   (a) The dispersion relations for a \ce{Bi2Te3} nanowire of radius $500$ \AA with magnetization  $\vec{M} = M (\cos(\phi_M), \sin(\phi_M), 0)$ for $\phi_M = 0$ (red lines), $\frac{\pi}{12}$ (green lines) and $\frac{\pi}{6}$ (blue lines). (b) The energy of the first band whose band bottom is visible in panel (a) plotted as a function of $k_z$ and $\phi_M$.  } 
\label{gT2combMuX2}
\end{figure}		

Differing from the TI nanocylinders studied in Ref. [\onlinecite{JAP117_17C749}],  the hexagonal warping terms in \ce{Bi2Te3} result in a dependence of the dispersion relations on the direction of the applied magnetization perpendicular to the cylinder axis. Panel (a) of Fig. \ref{gT2combMuX2} shows that for a given band, the value of $\phi_M$ which gives the lowest energy depends on both the band as well as the value of $k_z$. Panel (b) in turn shows that the energy of each band varies periodically with a period of $\pi / 3$ as $\phi_M$ is varied. This is reminiscent of the behavior of Fig. 2a of Ref. [\onlinecite{SciRep4_5062}]  in which the bandgap in the flat $+z$ terminated \ce{Bi2Te3} slab subjected to an in-plane magnetization exhibits a similar 6 fold periodic variation with the in-plane magnetization angle.   

\begin{figure}[ht!]
\centering
\includegraphics[scale=0.48]{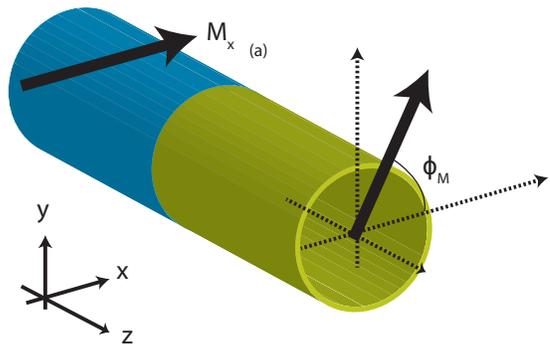}
\caption{   The system considered in this section consisting of a source \ce{Bi2Te3} cylinder magnetized in the $+x$ direction and a coaxial drain \ce{Bi2Te3} cylinder, both of radius $50 \mathrm{nm}$, coaxial with each other.    }
\label{gT2mr} 
\end{figure}

\begin{figure}[ht!]
\centering
\includegraphics[scale=0.33]{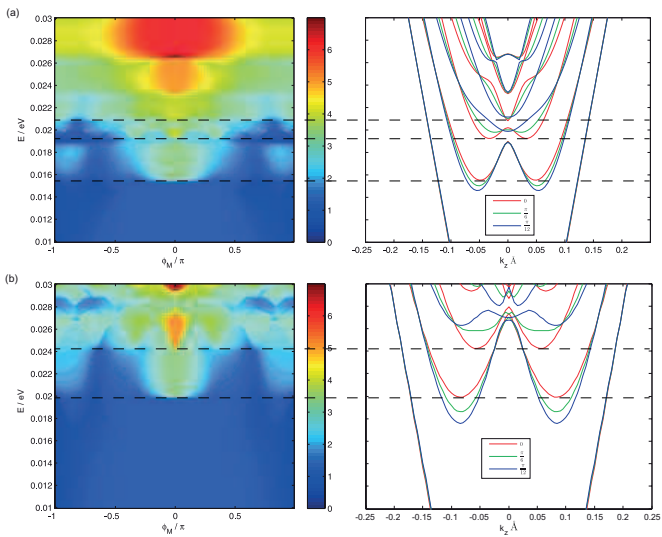}
\caption{   The left panels show transmission from a $50\ \mathrm{nm}$ TI cylinder magnetized in the $+x$ direction to a co-axial drain cylinder magnetized to the same magnitude and in the $(\cos(\phi_M),\sin(\phi_M),0)$ direction at (a) $M=0.02\ \mathrm{eV}\hbar^{-1}$ and (b) $M=0.03\ \mathrm{eV}\hbar^{-1}$ as a function of energy and drain magnetization angle $\phi_M$. The panels on the right show the dispersion relations at various magnetization angles at the same magnetization magnitude as the accompanying plots on the left, as a function of the wave vector $k_z$ and the energy. The energy values of the transmission profiles (left) and dispersion relations (right) are aligned to facilitate comparison. We have also drawn lines linking the energy at which key features in the transmission profile emerge with their corresponding features in the band diagrams.    }
\label{gcombMrX} 
\end{figure}

The combined effects of the band bending at small $|k_z|$ and the magnetization angle dependence give rise to a rich transmission profile as the energy and magnetization angles are varied.  Fig. \ref{gcombMrX} shows the transmission from a  source cylinder magnetized in the $+x$ direction to a coaxial drain cylinder  magnetized to the same magnitude shown in Fig. \ref{gT2mr}, as a function of the drain magnetization angle and the energy for two values of source magnetization.

We first consider panel (a) of Fig. \ref{gcombMrX} which corresponds to the smaller magnetization magnitude.  At energies below slightly more than $0.015\ \mathrm{eV}$, theew is only one propagating mode in both the source and drain. The transport in this regime exhibits the conventional magnetoresistance where the transmission is highest for parallel magnetizations between the source and drain, and lowest for the antiparallel configuration.  At energies lying between slightly more than $0.015\ \mathrm{eV}$ to about $0.025\ \mathrm{eV}$ the transmission exhibits the anomalous magnetoresistance in Ref. [\onlinecite{JAP117_17C749}]. The band-bending near  $|k_z=0|$ results in the transmission at antiparallel configuration being higher than at intermediate configurations between parallel and anti-parallel source and drain magnetization configurations. Brieftly, the anomalously high transmission at antiparallel transmission occurs because for a given direction of current flow, the states whose bands bend for a given magnetization direction, are localized at the same angular positions around the cylinder circumference as those states whose bands do not bend for the opposite magnetization direction. The overlap between the angular postions of where the wavefunctions have the largest ampltidue of the bent bands on one side, and the unbent bands on the other side, of the source-drain interface lead to increased transmission from the source to the drain and hence the anomalously high transmission \cite{JAP117_17C749}.  In contrast, at higher energies above approximately $0.023\ \mathrm{eV}$ in panel (a) where band bending does not occur we have again the conventional magnetoresistance.

A `sawtooth' profile of high transmission at $\phi_M = n \pi/3$  appears in a narrow energy range between about $0.019\ \mathrm{eV}$ to slightly less than $0.021\ \mathrm{eV}$. The emergence of this sawtooth pattern is due to the variation of the energy of the band bottom of the band emerging at this energy range with the drain magnetization direction -- the narrow tip of each  high transmission `tooth' at each integer $n$ $\phi_M = n\pi/3$ corresponds to the drain magnetization direction at which the band bottom has the lowest value of energy. At higher values of energy the band bottoms of the bands which subsequently emerge do not vary as much with the drain magnetization direction, and the bands no longer bend upwards at small $|k_z|$. This leads to a conventional magnetoresistance behavior at high energies. Comparing now between panels (b) and (a), we see that the larger magnetization  in (b) results in more extensive bending of the lower energy bands at small $|k_z|$. This results in a larger energy range over which the anomalous magnetoresistance appears. The larger variation of the energy of each band with the magnetization angle also leads to a larger energy range for the sawtooth transmission profile.  

The sawtooth transmission profile may possibly be of utility as a four-state magnetic memory element. The magnetic memory element may be biased and gated so that the total current is given by the energy integral over an energy range falling within the sawtooth transmission profile regime (about $0.024\ \mathrm{eV}$ to $0.027\ \mathrm{eV}$ ), and the 4 states represented by the drain magnetization lying in $\phi_M = 0$, $\pi/6$, $2\pi/6$ and $3\pi/6$. The sawtooth transmission profile then leads to a relatively large difference in current between memory states represented by  adjacent values of the drain magnetization directions  (e.g. between $\phi_M=\pi/6$ and $\phi_M=2\pi/6$) while retaining a current difference large enough between the $\phi_M=0$ and $\phi_M=2\pi/6$, and $\phi_M =\pi/6$ and $\phi_M=3\pi/6$ pairs, to be able to robustly distinguish between the four states.

\section{Conclusion}
In this work we derived the two-band effective Hamiltonian for the surface states of a \ce{Bi2Te3} nanocylinder incorporating the hexagonal warping terms. We calculated the dispersion relations from the underlying four-band Hamiltonian and saw that the energies at which the bulk states emerge is only weakly dependent on the cylinder diameter. The dispersion relations calculated from the two-band Hamiltonian match those of the four-band Hamiltonian reasonably well. The dependence of the \ce{Bi2Te3} dispersion relation on the angle at which the magnetization is applied perpendicular to the cylinder axis gives a rich transmission profile between two cylinders magnetized in different directions. In particular, the sawtooth profile which exists at some values of energy and large magnetizations may be of utility in a four-state magnetic memory bit.  

\section*{Acknowledgment}
The authors acknowledge the Singapore National Research Foundation for support under NRF Award Nos. NRF-CRP9-2011-01 and NRF-CRP12-2013-01, and MOE under Grant No. R263000B10112.

\section*{Appendix} 
Here, we briefly outline the derivation of the effective Hamiltonian for the surface states of a flat \ce{Bi2Te3} slab of infinite dimensions in the $x$ and $y$ directions, semi-infinite thickness in the $z$ direction, and terminated by a flat surface with the outward normal in the $+z$ direction. For convenience, we set the surface to be at $z=0$, so that the surface states which decay exponentially into the surface as $z\rightarrow-\infty$ have the form $\exp(\lambda^{(z)} z)$. 

In this case, the perpendicular Hamiltonian containing derivatives in the $z$ direction reads, in Cartesian coordinates, as
\begin{equation}
	\langle \vec{r}|H^{(z)}_{(4B),\perp} = M^{(z)}_\perp(\mathbb{I}_\sigma\otimes\tau_3) + B (\mathbb{I}_\sigma\otimes\tau_2) k_z \label{HzPt},
\end{equation}
where $M^{(z)}_\perp \equiv (M_0 - M_2\partial_z^2)$, and the $(z)$ superscripts indicate that these quantities pertain to the $+z$ terminated slab. We solve for the zero-energy eigenstates of $H^{(z)}_{(4B),\perp}$. These zero-energy eigenstates which are proportional to $\exp(\lambda^{(z)} z)$ and hence, consistent with our boundary condition, exist only when $M^{(z)}_\perp = -\lambda^{(z)} /B$ (similar to the case of the cylindrical geometry considered in the main paper). Solving $M^{(z)}_\perp = -\lambda^{(z)} /B$ for $\lambda^{(z)}$ in turn gives two solutions, i.e., $\lambda^{(z)}_{\pm} = (B \pm \sqrt{B^2+4M_0M_1})/(2M_1)$. The zero-energy eigenstates of Eq. \eqref{HzPt} can then be chosen to be proportional to 
\begin{eqnarray}
	\langle \vec{r} |+^{(z)}\rangle &\propto&(\exp(\lambda^{(z)}_+z) - \exp(\lambda^{(z)}_-z) )   \begin{pmatrix} 1 \\ -1 \\ 0 \\ 0 \end{pmatrix}, \\
	\langle \vec{r}  |-^{(z)}\rangle &\propto& (\exp(\lambda^{(z)}_+z) - \exp(\lambda^{(z)}_-z) ) \begin{pmatrix} 0 \\ 0 \\ 1 \\ -1 \end{pmatrix}.
\end{eqnarray} 

Taking the projection of the remaining terms of Eq. \eqref{ham} not already included in  Eq. \eqref{HzPt}, onto the $|\pm^{(z)}\rangle$ states then yields Fu's surface Hamiltonian of Ref. [\onlinecite{PRL103_266801}]
\begin{equation}
H = A (k_x\sigma_y - k_y\sigma_x) + R_1 (k_x^3 - 3 k_x k_y^2) \sigma_z,
\end{equation} 
upon identifying $A \rightarrow v_f$ and $R_1 \rightarrow \lambda $.

\end{document}